\documentclass[12pt]{iopart}
\usepackage{iopams}
\usepackage{cite}
\usepackage{graphicx}

\begin{document}

\title[Long-range proximity effect in Nb-based heterostructures]{Long-range proximity effect in Nb-based heterostructures induced by a magnetically inhomogeneous Permalloy layer}

\author{C Cirillo$^1$, S Voltan$^2$, E A Ilyina$^{1}$\footnote{Present address: CERN, Geneva, Switzerland.}, J M Hern\'{a}ndez$^{3,4}$, A Garc\'{i}a-Santiago$^{3,4}$, J Aarts$^2$ and C Attanasio$^{1}$}

\address{$^1$CNR-SPIN
Salerno and Dipartimento di Fisica \lq\lq E. R. Caianiello\rq\rq,
Universit\`{a} degli Studi di Salerno, via Giovanni Paolo II 132, I-84084 Fisciano (Sa), Italy}
\address{$^2$Kamerlingh Onnes-Huygens Laboratory, Leiden University,
P.O. Box 9504, 2300 RA Leiden, The Netherlands}
\address{$^3$Grup de Magnetisme, Departament de F\'{i}sica de la Mat\`{e}ria Condensada,
Facultat de F\'{i}sica, Universitat de Barcelona, c.\,Mart\'{i} i
Franqu\`{e}s 1, planta 4, edifici nou, 08028 Barcelona, Spain}
\address{$^4$Institut de Nanoci\`{e}ncia i Nanotecnologia IN2UB, Universitat de Barcelona, 08028 Barcelona, Spain}
\ead{attanasio@sa.infn.it}
\vspace{10pt}
\begin{indented}
\item[]
\end{indented}

\begin{abstract}
Odd-frequency triplet Cooper pairs are believed to be the carriers of long-range superconducting correlations in ferromagnets. Such triplet pairs are generated by an inhomogeneous magnetisation at the interface between a superconductor (S) and a ferromagnet (F). So far, reproducible long-range effects were reported only in complex layered structures designed to provide the magnetic inhomogeneity. Here we show that spin triplet pair formation can be found in simple unstructured Nb/Permalloy ($\rm Py=Ni_{0.8}Fe_{0.2}$)/Nb trilayers and Nb/Py bilayers, but only when the thickness of the ferromagnetic layer ranges between 140 and 250 nm. The effect is related to the emergence of an intrinsically inhomogeneous magnetic state, which is a precursor of the well-known stripe regime in Py that in our samples sets in at thickness larger than 300 nm.
\end{abstract}
\noindent{\it Keywords\/}: superconductivity, magnetism, heterostructures, proximity effect, dimensional crossover

%
%
%
%
%

\section{Introduction}
Superconductivity and ferromagnetism are competing phases whose coexistence is unlikely to occur. Notable exceptions take place when the electrons responsible for the magnetism are only weakly coupled to those inducing superconductivity, as in some ternary rare-earth compounds \cite{Fertig}. Differently from the case of bulk systems, the coexistence between superconductivity and ferromagnetism may be easily achieved in artificial superconductor-ferromagnet (S/F) hybrids. In these systems the two antagonistic orderings are confined in spatially separated layers interacting via the proximity effect, which arises when a superconductor comes in metallic contact with a ferromagnet \cite{Buzdin}. In this case, the spin-singlet Cooper pairs enter the F-layer and magnetic excitations leak into the S-region across the S/F interface. As confirmed by many experiments \cite{Lazar,Cirillo}, the penetration depth, $\xi_{\rm F}$, of singlets in the F-layer is, in the diffusive limit, basically given by $\xi_{\rm F} =\sqrt{\hbar D_{\rm F}/E_{\rm ex}}$ ($D_{\rm F}$ is the diffusion coefficient), while superconductivity is suppressed in S within a distance $\xi_{\rm S}$ from the interface ($\xi_{\rm S}$ is the superconducting coherence length). In addition, the presence of the exchange field, $E_{\rm ex}$, in F causes an energy shift between the electrons of the pairs entering the F-layer and the Cooper pairs acquire a non-zero center-of-mass momentum. As a consequence, the superconducting order parameter does not decay monotonically in F, as it would happen in the case of a normal metal, but it shows oscillatory decay in the direction perpendicular to the interface over a length scale given (again) by $\xi_{\rm F}$ \cite{Ryazanov,Robinson}. In strong ferromagnets, since $E_{\rm ex} \sim 1$ eV, $\xi_{\rm F}$ is only few nanometers. 

However, at the interface between a superconductor and a ferromagnet, conventional singlet Cooper pairs can be converted into equal-spin triplet ones. Since the triplets have their spins equally aligned, they are much less affected by the pair breaking caused by $E_{\rm ex}$ in F. Thus, once injected in the F-layer, at low $T$ they can survive over distances of the order of  hundreds of nanometers \cite{Bergeret}, contrary to what happens for the singlets. Such spin-triplet correlations are predicted to have even symmetry in space (s-wave), which makes them robust against scattering, but have odd symmetry with respect to time (hence named odd-frequency). The key factor to achieve singlet-to-triplet conversion is the presence of a certain degree of magnetic inhomogeneity at the S/F interface \cite{Bergeret}. There are different ways of providing such inhomogeneity. In the original theoretical proposal, the magnetic inhomogeneity was described in terms of a rotating vector with the angle of the magnetization direction rotating in the plane of the S/F interface when moving away from it. This scenario could be realized, for example, in a domain wall within which the magnetization gradually rotates. So far, however, almost all the experiments which gave evidence of a long-range proximity effect relied on different ways to provide the required magnetic inhomogeneity \cite{Keizer,Khaire,Sprungmann,Anwar}. In most of the experimental works, an extra ferromagnetic layer F$_1$ is inserted in between S/F and the inhomogeneity is controlled by varying the collinearity between the magnetization of F and F$_1$\cite{Robinson1,Leksin,Zdravkov,Banerjee,Flokstra,Singh}. By using holmium (Ho) as F$_1$-layer and Co as F-layer, Robinson \etal \cite{Robinson1} more closely reproduced the original theoretical model. Ho, indeed, is a rare-earth ferromagnet with conical magnetic ordering, whose magnetization vector rotates around the $c$-axis, if one moves along it. In this case, the inhomogeneity is expected to be intrinsically present in the Ho layer, however the multilayer geometry F$_1$/F/F$_1$ is still needed. For these heterostructures a theoretical explanation is also available \cite{Fritsch1,Fritsch2}. 

In this framework the properties of Permalloy can be particularly useful since it is well known that, if grown under certain specific conditions, it can form stripe-domains \cite{Saito,Amos,Belkin}. This is realized when its thickness exceeds a critical value, $d_{\rm cr}$, which depends on the growth parameters such as, among others, the deposition rate and the substrate temperature \cite{Amos,BenYoussef,Dastagir}. In this configuration the magnetization vector lies mainly in-plane, parallel to the stripe direction in all domains, but it develops an out-of-plane component which goes alternately upward and downward. The out-of-plane component of the different stripes is therefore aligned antiparallel with Ne\'{e}l-domain-walls in between, in which the magnetization rotates coherently. Recently, we extensively described the magnetic properties of Py and we characterized the stripe-domain regime \cite{Voltan}. We also showed that below $d_{\rm cr}$ exists a broad regime, approximately between 0.5 $d_{\rm cr}$ and $d_{\rm cr}$, where the magnetization can easily become inhomogeneous without being arranged in stripes. We called this state emerging stripe-domain regime. The intrinsic magnetic inhomogeneity of Py implied in the occurrence of a stripe-domain phase led us to investigate the possibility of using it as possible generator for triplet correlations. The question is whether it is possible to have S/F/S (or S/F) structures where the conversion is intrinsically provided by the F-layer itself, due to its magnetic configuration, as proposed in the reference \cite{Bergeret}.

In this article we investigate the temperature dependence of the parallel upper critical field, $H_{\rm c2 \parallel}(T)$, of simple Nb/Py/Nb trilayers and Nb/Py bilayers. The thickness of the Nb layers, $d_{\rm Nb}$, is kept constant at 25 nm while the thickness of the Py layer, $d_{\rm Py}$, is varied across the different thickness regimes: homogeneous (H), emerging stripe-domain (ESD) and stripe-domain (SD). For the trilayer with $d_{\rm Py}$ in the ESD regime, namely for 125 nm $\lesssim d_{\rm Py} \lesssim$ 300 nm, a 2D--3D dimensional crossover (DCO) was observed at  $T \simeq 0.9\,\,T_{\rm c}$, where $T_{\rm c}$ is the superconducting critical temperature of the system. Moreover, a clear kink is present in the $H_{\rm c2\parallel}(T)$ curves of Nb/Py bilayers when $d_{\rm Py}=200$ nm. These observations, which we attribute to an increased effective thickness of the superconducting layer, cannot be explained within the spin-singlet proximity effect, because of the short coherence length of Py, estimated to be about 1.9 nm \cite{Robinson,Moraru}. The results are rather compatible with a long-range spin-triplet proximity effect, induced by the inhomogeneous magnetic configuration of the ESD regime.

\section{Experimental methods}

Nb/Py/Nb trilayers and Nb/Py bilayers were grown on Si(100) substrates by ultrahigh vacuum dc diode magnetron sputtering at an Ar pressure of $2.25 \times 10^{-3}$ Torr after obtaining a base pressure of $1.5 \times 10^{-8}$ Torr. The substrates were nominally kept at room temperature during the deposition process. The typical deposition rates were 0.25 nm/s for Nb and 0.30 nm/s for Py measured by a quartz crystal monitor previously calibrated by low-angle x-ray reflectivity measurements on deliberately deposited thin films of each material. The prepared samples are unstructured and the typical in-plane dimensions are $5 \times 10 \,\,\, {\rm mm}^2$. Samples have constant Nb thickness, $d_{\rm Nb}=25$ nm, and variable Py thickness with $d_{\rm Py}$ in the range $20-430$ nm. Thanks to the presence of a movable shutter in the deposition chamber, which selectively covers the substrates, three different samples can be grown in the same deposition run. A single 25-nm-thick Nb film and several single Py films, having the same thickness as the Py layers in the corresponding hybrids, were also deposited and characterized for comparison.

Single layers of Py were magnetically characterized in-depth using Magnetic Force Microscopy (MFM), ferromagnetic resonance (FMR), SQUID magnetometry and magnetoresistance measurements (MR). Details of the used techniques can be found in the reference \cite{Voltan}.

The resistive transitions of unstructured multilayers were performed using an in-line four-terminal geometry with a constant bias current of 500 $\mu$A. The distance between the current (voltage) pads was about 8 mm (3 mm). The samples, mounted on a copper block and placed at the center of a NbTi superconducting solenoid, were immersed in a $^4$He cryostat. During measurements the temperature stabilization was around 1 mK.

\begin{figure}[t!]
\centering%
\includegraphics[scale=0.4]{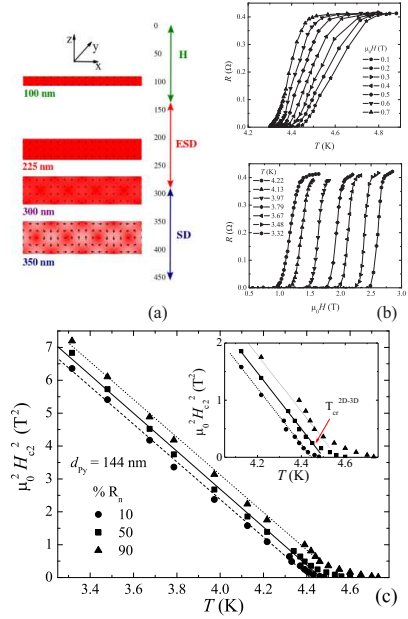}%
\vskip 2truecm
\caption{{\small (a) Cross sectional magnetization distribution simulated by OOMMF for four different values of $d_{\rm Py}$,
namely $d_{\rm Py} = 100, 225, 300, 350$ nm, obtained using the parameters presented in the reference \cite{Voltan}. The
samples are considered to be infinite along the $y$-axis (2D simulations) with lateral dimension
(along $x$) much larger than the region of interest $\rm (6\, \mu m)$. The arrows schematically indicate the
direction of the magnetization component in the $xz$-plane, while the color shows whether the
magnetization vector is parallel to the $y$-axis (red) or it deviates from it (white). The first indication
of inhomogeneities start to be visible for $d_{\rm Py} = 300$ nm (see arrows); for $d_{\rm Py} = 350$ nm the stripe domains
are developed. (b)
 Resistive transitions and superconducting phase diagram for the trilayer Nb(25)/Py(144)/Nb(25). Top panel: $R(T)$ curves at different
values of the in-plane applied magnetic field. Bottom panel: Resistance $R$ as a function of the in-plane
field at different temperatures. (c) Temperature dependence of $H_{\rm c2 \parallel}^{2}$ for the trilayer Nb(25)/Py(144)/Nb(25) determined using three different resistive criteria. The crossover
temperature can be easily estimated as the point where the linear fit, which in a quadratic scale
identifies the 2D regime, deviates from the experimental data (see inset of the figure).}} \label{Magnetic characterization}
\end{figure}

\section{Results}
\subsection{Magnetic characterization}
A detailed magnetic characterization of Py as a function of the layer thickness was
recently presented in the reference \cite{Voltan}. MFM measurements on unstructured Py single layers \cite{Voltan} showed that magnetic stripe-domains are visible only above a thickness $d_{\rm Py} \simeq 300$ nm, which we defined to be the critical thickness value $d_{\rm cr}$, i.\,e.\,\,the
lower boundary of the stripe-domain regime (SD). For $d < d_{\rm cr}$, two different regimes
could be recognized: a homogeneous regime (H), for $d \lesssim 125$ nm, and a so-called
emerging stripe-domain regime (ESD), for 125 nm $\lesssim d \lesssim $ 300 nm. For the samples in
the ESD regime, the MFM measurements did not provide evidence of an inhomogeneous
magnetic configuration, except for thicknesses close to the transitions to the SD
regime. This suggests that either the magnetization is fully in-plane or the out-of-plane component is too weak to be 
detected by the MFM technique \cite{Voltan}. However, signatures of inhomogeneity emerged with other measurement techniques. Magnetic hysteresis loops $M(H)$ for samples in the ESD regime showed
hints of a linear dependence before the saturation was reached. This linear behavior in general is a well-known feature which signals the presence of stripes-domains and it is attributed to the coherent rotation of the stripes before the saturation. The observation
of such dependence in the ESD regime, although less pronounced, is an indication
of a certain degree of inhomogeneity, even if not necessarily in the stripe-form. The
tendency of the samples in the ESD regime to have an inhomogeneous configuration
emerged more clearly by looking at structured samples \cite{Voltan}, in particular in the domain-wall
configuration and in the outcome of MR measurements. The domain wall
configuration of the structured samples was reproduced by simulations realized
with the object oriented micromagnetic framework (OOMMF) software for all the three
different regimes \cite{Voltan}.

Since the devices we studied here are unstructured, we performed
simulations on semi-infinite samples, for different thicknesses. The results are shown
in figure \ref{Magnetic characterization}(a). The magnetic parameters used (exchange stiffness constant, $A$, saturation
magnetization, $M_{\rm s}$, and out-of-plane anisotropy, $K_{\rm \perp}$) are the same as in the reference \cite{Voltan}, namely $A= 13 \times 10^{-12}$ J/m, $M_{\rm s}=8.59 \times 10^5$ A/m, and $K_{\rm \perp}=5.6 \times 10^3$ J/m$^3$. The sample
is considered to be infinite along the $y$- and $x$-axes while it spans the thickness of the
sample along the $z$-axis. The cross section shown in the figure is thus taken in the $xz$-plane,
while the $y$-axis points inside this plane and represents the direction along which
the sample is initially magnetized, prior to the magnetic measurements. The color code
indicates the direction of the magnetization with respect to the $y$-axis, namely red and
white mean the magnetization is respectively parallel or perpendicular to the initial direction.
Black arrows indicate the orientation of the components of the magnetization
in the $xz$-plane. The magnetization in the thinner samples ($d_{\rm Py}=100$, 225 nm
topmost sketches) stays parallel to the initial state (full dark color, indicating an homogeneous
magnetic state), while deviations from such state begin to occur at $d_{\rm Py}=300$
nm, for which areas with magnetization perpendicular to the $y$-axis start to appear. As
$d_{\rm Py}$ increases, the size and density of such areas increase and the magnitude of the component
of the magnetization which deviates from the $y$-axis also grows. At $d_{\rm Py}=350$ nm
(bottom sketch), the domains with an out-of-plane magnetization component become
wider, with thinner domain walls in between. This is the SD regime. Thus, from these simulations no clear evidence of inhomogeneity appear for the ESD regime. However,
it is possible that the \lq\lq semi-infinite\rq\rq\ approximation only partially reproduces the real
physics of our devices. Furthermore, in the simulations the role of the proximity with
superconducting layer is not taken into account. In the study below, the magnetic field
is applied parallel to the film plane, and consequently the magnetic flux perpendicular
to interface is minimum, however the diamagnetic nature of Nb below $T_{\rm c}$ could influence
the magnetic configuration at the interface.

\subsection{Superconducting transport properties of Nb/Py/Nb trilayers}
In order to determine the  $H_{\rm c2}(T)$ phase diagrams of the Nb/Py/Nb trilayers, the resistance $R$ was measured either as a function of the temperature $T$ (at a fixed applied magnetic field $H$) or as a function of $H$ (at a fixed $T$). The field was applied in-plane, perpendicular to the direction of the bias current. In figure \ref{Magnetic characterization}(b) a selection of $R(T)$ and $R(H)$ curves for the trilayer Nb(25)/Py(144)/Nb(25) is presented (numbers in parentheses indicate the thickness expressed in nanometers). $T_{\rm c}$ was defined at $T_{50\%}$, namely at the temperature at which the resistance value is 50\% of the normal state resistance $R_{\rm N}$, measured at $T = 10$ K. Before measuring, a strong magnetic field (approximately 1 T) was applied in the plane of the substrate at low temperatures and then removed. This was done in order to \lq\lq induce\rq\rq\, the stripes in the SD regime, and for consistency in the other two regimes. The width of the transitions at zero field, defined by $T_{90\%}-T_{10\%}$, is about 200 mK for all the samples and does not increase when a field is applied. The single Nb film 25 nm-thick has a critical temperature (at zero field) $T_{\rm c}=6.5$ K and shows a two-dimensional (2D) behavior \cite{Chun} ($H_{\rm c2\parallel}(T)\propto \sqrt{1-T/T_{\rm c}}$) in the whole investigated temperature range. The Ginzburg-Landau (GL) coherence length at zero temperature, $\xi_{\rm GL}(0)$, was extracted from the linear temperature dependence of the perpendicular upper critical field $H_{\rm c2\perp}(T) = (\phi_0/2 \pi \xi^2_{\rm GL}(0)) (1-T/T_{\rm c})$ \cite{Schock}. The obtained value is $\xi_{\rm GL}(0) \simeq 10$ nm, implying a superconducting coherence length $\xi_{\rm S}(0)=(2/\pi)\xi_{\rm GL}(0)\simeq 7$ nm. 

\begin{figure}[t!]
\centering
\includegraphics[scale=0.4]{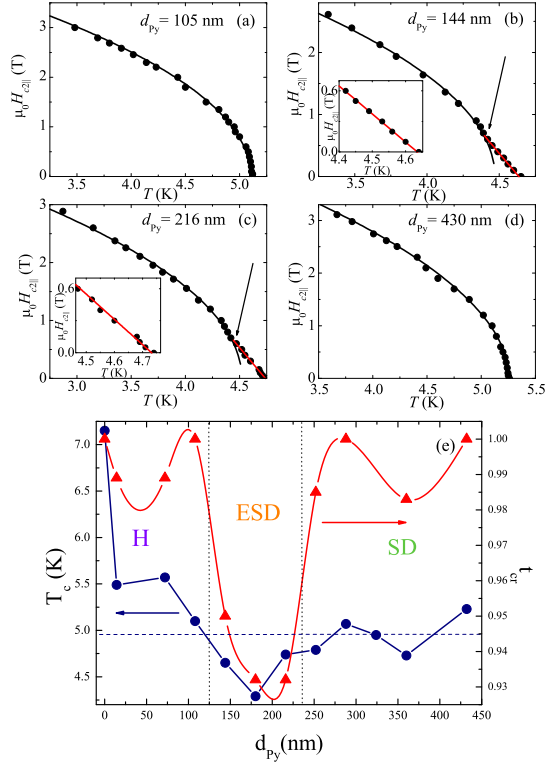} 
\vskip 1truecm
\caption{{\small $H_{\rm c2\parallel}(T)$ phase boundaries of Nb/Py/Nb trilayers with $d_{\rm Nb}=$ 25 nm and (a) $d_{\rm Py}=$ 105 nm, (b) $d_{\rm Py}=$ 144 nm, (c) $d_{\rm Py}=$ 216 nm,
and (d) $d_{\rm Py}=$ 430 nm. Black lines show the square-root (2D) temperature dependence of $H_{\rm c2\parallel}$, that is $H_{\rm c2\parallel}(T)=H_{\rm c2\parallel}(0) \sqrt{1-T/T_{\rm c}}$. Red lines show the linear (3D) temperature dependence of $H_{\rm c2\parallel}$, that is  $H_{\rm c2\parallel}(T)=H_{\rm c2\parallel}(0) (1-T/T_{\rm c})$. The curves were obtained using $H_{\rm c2\parallel}(0)$ as the only fit parameter. The arrows indicate $T_{\rm cr}^{2D-3D}$. Insets of (b) and (c): high-temperature region of the corresponding $H_{\rm c2\parallel}(T)$ phase diagrams. (e) $t_{\rm cr}^{2D-3D}=T_{\rm cr}^{2D-3D}/T_{\rm c}$ as obtained from $H_{\rm c2\parallel}(T)$ measurements (right scale) and $T_{\rm c}$ (left scale) as a function of $d_{\rm Py}$ in Nb/Py/Nb trilayers. The red and the blue lines are guides to the eye. The dashed vertical lines define the three different regions corresponding to the magnetic configuration (H, ESD, SD) of the Py films. According to the values of $t_{\rm cr}^{2D-3D}$, in the ESD interval the two outer Nb layers are coupled.}} \label{Hc2}
\end{figure}

The $H_{\rm c2\parallel}(T)$ phase diagrams for a representative set of the trilayers are presented in figure \ref{Hc2}. The samples Nb(25)/Py(105)/Nb(25) (figure \ref{Hc2}(a)) and Nb(25)/Py(430)/Nb(25) (figure \ref{Hc2}(d)) show a 2D like behavior in the whole temperature range. The black line is the square-root temperature dependence of $H_{\rm c2\parallel}(T)$ obtained leaving $H_{\rm c2\parallel}(0)$ as the only fitting parameter. The first Py thickness is in the H regime, the latter in the SD regime. In both cases, thus, the Nb layers behave as the isolated single layer. Like in any conventional S/F interface, the singlet component cannot penetrate the Py layer more than 1-2 nm (on both sides), so the Nb layers result isolated. In the ESD regime, instead, the behavior is quite different. For both Nb(25)/Py(144)/Nb(25) (figure \ref{Hc2}(b)) and Nb(25)/Py(216)/Nb(25) (figure \ref{Hc2}(c)) there is a 2D--3D dimensional crossover (DCO) at a temperature $T_{\rm cr}^{2D-3D} \simeq 0.9\,\,T_{\rm c}$. Close to $T_{\rm c}$, $H_{\rm c2\parallel}(T)$ is linear, as for a three-dimensional (3D) system, while at $T_{\rm cr}^{2D-3D}$ it presents a square-root behavior of a 2D system. The insets in these panels show an enlargement of the data for temperatures close to $T_{\rm c}$ where the linear behavior of $H_{\rm c2\parallel}(T)$ is much more evident. The DCO indicates a change in the dimensionality of the superconducting layers (in relation to $\xi_{\rm GL}$) and can be explained within the framework of a long-range proximity effect. The peculiar inhomogeneous magnetic configuration of the ESD layer generates equal-spin triplet Cooper pairs which can \lq\lq leak\rq\rq\ into Py, therefore extending the effective thickness of the superconducting layers.
One possible explanation is that the effect, as a consequence of the presence of magnetic inhomogeneities in Py, is due to a coupling between the top and bottom Nb layer across the Py. However, given the length scales involved, this seems to be very unlikely. The penetration length of the triplet component, indeed, is expected to be limited by the spin diffusion length, $\ell_{\rm sf}$, which for Py is relatively small, about 5 nm \cite{Moraru,Bass}, much shorter than the Py thickness. The reason why the 3D-behavior is observed only above $T_{\rm cr}^{2D-3D}$ has to do with the temperature dependence of the coherence length $\xi_{\rm S}(T) = \xi_{\rm S}(0)/\sqrt{1-T/T_c}$ and it will be more extensively discussed later. The previous analysis of the $H_{\rm c2\parallel}(T)$ phase diagrams is summarized in figure \ref{Hc2}(e) (right hand scale) where the reduced crossover temperature $t_{\rm cr}^{2D-3D}=T_{\rm cr}^{2D-3D}/T_{\rm c}$ is reported for all trilayers. The crossover temperature can be more easily estimated by plotting $H_{\rm c2\parallel}^{2}$ vs $T$, as shown in figure \ref{Magnetic characterization}(c) for the trilayer with $d_{\rm Py}=144$ nm: $T_{\rm cr}^{2D-3D}$ is the value at which the dependence deviates from the linear fit (see inset of the figure). For values of $d_{\rm Py}$ up to 125 nm (region H) and larger than 300 nm (region SD) $t_{\rm cr}^{2D-3D}$ is essentially equal to 1 (namely there is no crossover), while in the ESD region it is $t_{\rm cr}^{2D-3D}=0.93-0.95$. As a remark, as we show in figure \ref{Magnetic characterization}(c), the choice of the 50\% of $R_{\rm N}$ as a criterion to obtain the $H_{\rm c2\parallel}(T)$ phase boundaries of the different heterostructures does not alter the main results presented above, since the position of the 2D--3D crossover in the $H_{\rm c2\parallel}(T)$--plane, if present, is confirmed also if the 10\% or 90\% criteria for the determination of $T_{\rm c}$ are adopted. The dependence of $T_{\rm c}$ (at zero field) on the thickness of Py, plotted in figure \ref{Hc2}(e) (left scale), also shows a clear variation, with a dip in the middle of the ESD regime. Notably, $T_{\rm c}$ for the trilayer with $d_{\rm Py}=170$ nm is at least 0.5 K lower than $T_{\rm c}$ of the trilayers with $d_{\rm Py}$ above 300 nm. This is fully consistent with the explanation for the 2D--3D crossover. When triplets are formed, the leakage of Cooper pairs into the Py layer depletes the superconducting order parameter on the S side of the S/F interface. The length scale of the effect is determined by $\xi_{\rm S}(0)$ and since the thickness of the superconducting layer is only a few times $\xi_{\rm S}(0)$, the result is a suppression of $T_{\rm c}$ of the whole layer. This observation was predicted theoretically \cite{Fominov} and recently demonstrated in different systems \cite{Leksin,Singh}.

\subsection{Superconducting transport properties of Nb/Py bilayers}
\begin{figure}[t]
\centering
\includegraphics[scale=0.5]{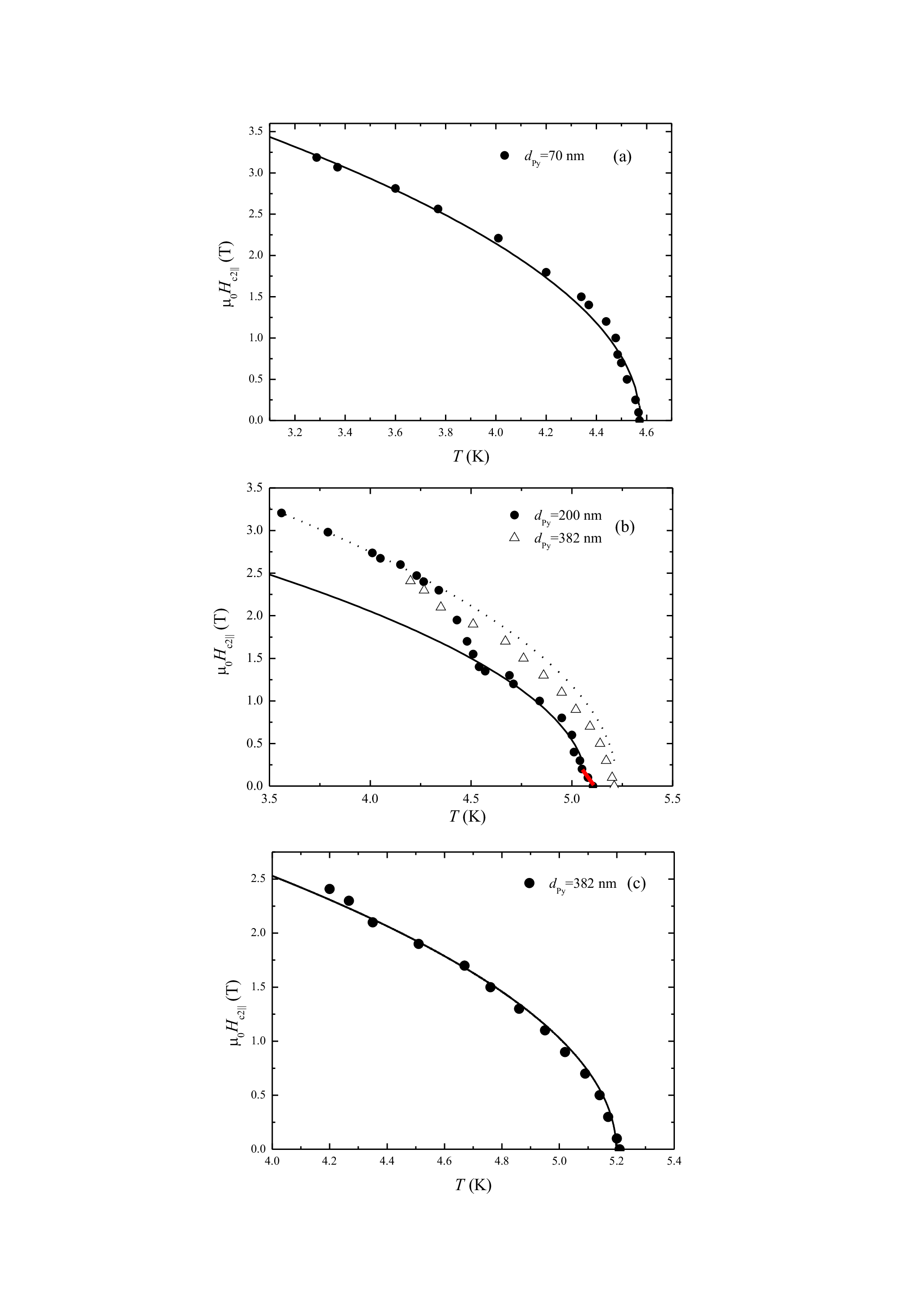}
\caption{{\small $H_{\rm c2\parallel}(T)$ phase boundaries of Nb/Py bilayers with $d_{\rm Nb}=$ 25 nm and (a) $d_{\rm Py}=$ 70 nm, (b) $d_{\rm Py}=$ 200 nm (closed circles) and  $d_{\rm Py}=$ 382 nm (open triangles), and (c) $d_{\rm Py}=$ 382 nm. Black lines in panels (a) and (c) show the 2D temperature dependence of $H_{\rm c2\parallel}$. Both curves were obtained using $H_{\rm c2\parallel}(0)$ as the only fit parameter. In panel (b) the red line shows the linear dependence of $H_{\rm c2\parallel}(T)$ near $T_{\rm c}$ while the solid (dashed) black line shows the square-root dependence of $H_{\rm c2\parallel}(T)$ for temperatures lower (higher) than $T=4.5$ K.}} \label{Hc2bilayers}
\end{figure}

Since the observed behavior is expected to be due to two separate S-layers at top and
bottom, decoupled by the thick Py layer, a similar effect should
be observed for a Nb/Py bilayer. In figure \ref{Hc2bilayers} we show the phase diagrams $H_{\rm c2\parallel}(T)$ for the bilayers Nb/Py, with $d_{\rm Nb}=25$ nm and $d_{\rm Py} = 70, 200, 382$ nm, in order to explore again the H, ESD, and SD regions, respectively. 

As expected, for  $d_{\rm Py} = 70$ and 382 nm
(figure \ref{Hc2bilayers}(a) and (c), respectively; H and SD regime) the dependence is 2D in the whole temperature range, as for the trilayers. The black lines are the fits of the 2D relation,
obtained using $H_{\rm c2\parallel}(0)$ as the only fitting parameter. The bilayer with $d_{\rm Py}=200$ nm (ESD regime), which in figure \ref{Hc2bilayers}(b) is compared with the Nb(25)/Py(382) one, also shows a DCO
but in this case the transition is 2D--2D, at about $T = 4.5$ K $(t_{\rm cr}^{2D-2D} \simeq 0.9)$. Very close to $T_{\rm c}$ there is a hint of 2D--3D dimensional crossover, but the small range makes it difficult to judge whether it is a real feature or an artifact. A possible explanation for the observations is given later.

\section{Discussion}

From the analysis of the phase diagrams of the S/F/S trilayers it emerges that a 2D--3D DCO is present only when the thickness of the Py layer is in the ESD regime.
Such a DCO is typically observed in S/N multilayers \cite{Chun} (with N a normal metal), and
it is ascribed to a change in the dimensionality of the superconducting layer with respect
to the coherence length $\xi_{\rm GL}(T)$. This occurs when the thickness, $d_{\rm S}$, of a single
layer is lower than (or of the order of) the coherence length but the total
thickness of two (or more) layers is larger. When $\xi_{\rm GL}$, which is temperature dependent,
becomes of the order of the spacer length $d_{\rm N}$, the S-layers are coupled and the $H_{\rm c2\parallel}(T)$ dependence becomes linear (3D--behavior), provided that $d_{\rm N} < 2 \xi_{\rm N}$. 
The latter condition makes a DCO unexpected in a S/F multilayer unless $d_{\rm F}$ is very small \cite{Koorevaar} or the ferromagnet is diluted \cite{Schock}. In our case $d_{\rm F} \gg \xi_{\rm F}$. For Py, indeed, $E_{\rm ex} \simeq 200$ meV \cite{Robinson} and the diffusion coefficient can be obtained via the relation $D_{\rm F}=(1/3)v_{\rm F} \ell_{\rm F}$, where  $v_{\rm F}=2.2 \times 10^5$ m/s \cite{Robinson} is the Fermi velocity and $\ell_{\rm F}$ is the mean free path. $\ell_{\rm F}$ can be obtained from $\rho\ell_{\rm F} = 31.5 \times 10^{-6}$ $\mu \Omega$ cm$^2$ \cite{Mayadas}, knowing that $\rho=20$ $\mu \Omega$ cm (value measured at low temperatures in our samples). Thus, from $\xi_{\rm F} =\sqrt{\hbar D_{\rm F}/E_{\rm ex}}$ , we obtain $\xi_{\rm F} \simeq 1.9$ nm. The scenario is different if long-range equal-spin triplet Cooper pairs are induced, because of the inhomogeneity in the ESD regime. In this case the coherence length is given by $\xi_{\rm F}^{\rm T}(T) = \sqrt{\hbar D_{\rm F}/2 \pi k_{\rm B} T}$ \cite{Bergeret}. By using the value of $D_{\rm F}$ extrapolated above, at $T=4.2$ K, $\xi_{\rm F}^{\rm T}(T=4.2$ K$) \simeq 20$ nm $\gg \xi_{\rm F}$. While estimating the length scale of the proximity effect, also the spin diffusion length $\ell_{\rm sf}$, typically the main limiting factor, has to be taken into account. For strong ferromagnets such as Co, $\ell_{\rm sf}$ is about 60 nm \cite{Bass,Piraux}, while for Py is expected to be much shorter, $\ell_{\rm sf} \simeq 5$ nm \cite{Moraru,Bass}. Therefore, the possibility that the two Nb layers are coupled across a 200 nm-thick Py layer is extremely unlikely even tough one can consider that for $d_{\rm Py}=144$ nm the order parameter is attenuated at the center of the ferromagnetic layer by the factor $\exp(-d_{\rm Py}/2\ell_{\rm sf}) = 5.5 \times 10^{-7}$, value that is much larger than $\exp(-d_{\rm Py}/2\xi_{\rm F}) = 3.5 \times 10^{-17}$, as expected in the spin-singlet scenario. The DCO in this case seems to be the result of the extended effective S-thickness, with only the proximity of the top S/F interface contributing. A similar DCO, indeed, was observed for a simple Nb/Cu bilayer \cite{Otop}. Whether the estimated short penetration length, due to $\ell_{\rm sf}$, is enough to explain a transition from a 2D to a 3D regime, is unclear. The spin diffusion length for Py was evaluated with a two-current model \cite{Dubois}, considering a homogeneously magnetized layer of Py. In the ESD regime, however, there is a certain degree of inhomogeneity which could be responsible for mixing the two channels, resulting in an enhanced effective spin-diffusion length \cite{Ruediger,Kent}. Moreover, the observation of a dip in the $T_{\rm c}$ vs $d_{\rm Py}$ in correspondence of the ESD regime is a further strong indication of the origin of the effects being the leakage triplet Cooper pairs into Py. If the suppression of $T_{\rm c}$ in the trilayer was only due to the leakage of spin-singlet Cooper pairs, the critical temperature values in the studied Py thickness range should be $d_{\rm Py}$ independent, because in this case $d_{\rm Py} \gg \xi_{\rm F}$. The reason why the triplet generation occurs only in the ESD regime has to do with the particular inhomogeneity of the magnetization in this thickness region. One could expect that the stripe-domains in the SD regime, and in particular the rotating domain-walls between stripes, could be a source of singlet-to-triplet conversion. However, this is not what the outcome of the measurements suggests. This can be explained by looking at the length scale of the inhomogeneities in this case. The typical width $w$ of the stripes is of the same order of magnitude of the thickness $d_{\rm Py}$ \cite{Hubert}, thus $w > 300$ nm, with the domain-wall width $\delta_{w}$ of a similar order of magnitude ($\delta_{w} \approx w/3$). This length scale is large, if compared to $\xi_{\rm F}$. As a consequence the short-ranged Cooper pairs injected into Py \lq\lq feel\rq\rq\ an homogeneous magnetization and therefore there is no conversion. In the ESD regime, instead, the magnetization is not arranged in stripes. Because of the weak perpendicular anisotropy the magnetic moments have a \lq\lq tendency\rq\rq\ to rotate out-of-plane, and the result is the presence of more localized inhomogeneities \cite{Hubert}. The 3D--2D transition which occurs by lowering the temperature and increasing the magnetic field, is probably due
to the alignment of the magnetization which is no longer inhomogeneous. Surprisingly, the transition field ($\simeq 0.5$ T) is an order of magnitude higher than the field of complete saturation \cite{Voltan}. A possible explanation could be that the uniaxial anisotropy and 
the interface roughness lead to a residue of unsaturated and misaligned moments close to the interface. Triplet correlations can 
actually be induced by quite small amounts of misaligned moments, as already noticed in early work involving CrO$_2$: in sandwiches of 
NbTiN/CrO$_2$ \cite{Keizer} and MoGe/CrO$_2$ \cite{Anwar} triplet correlations were found in the absence of engineered magnetic inhomogeneities and 
they were sustained in fields well above the nominal saturation field of CrO$_2$.

The measurements performed on the bilayers confirm the picture described above, with a DCO observed only for the ESD regime, and at an even higher magnetic field
($\simeq 1.5$ T). Why the transition in this case is 2D--2D is not entirely trivial. In the reference \cite{Neerinck} a 2D--2D DCO was observed in a Pb/Ge multilayer because the total thickness of the coupled Pb layers are still in the 2D regime. Here, if we fit the $H_{\rm c2\parallel}(T)$ curve of Nb(25)/Py(200) at low temperatures, which coincides with the curve of Nb(25)/Py(382) (figure \ref{Hc2bilayers}) using the expression $\mu_0 H_{\rm c2\parallel}(0)=(\sqrt{12} \phi_0) /(2 \pi \xi_{\rm GL}(0) d_{\rm S_{eff}})$ \cite{Jin}, we obtain an effective superconducting thickness $d_{\rm S_{eff}}$ of about 20 nm. This is also the thickness inferred from the Nb(25)/Py(382) data. A lower thickness than the actual layer thickness is to be expected
for an S/F bilayer where strong pair breaking on the F-side of the interface lowers the
Cooper pair density. From the extrapolation close to $T_{\rm c}$, instead, the estimated $d_{\rm S_{eff}}$ value is approximately 25 nm. On the one hand, this is significantly larger than the low-temperature thickness;
on the other hand, it is thin enough to be in the 2D regime. Maybe even more importantly,
the crossover in the $H_{\rm c2\parallel}(T)$ curve of Nb(25)/Py(200) suggests that not only the effective thickness decreases, but that also $T_{\rm c}$ increases, which would be in full agreement with the triplet leakage
scenario. An effective thickness of 25 nm could even be expected to arise from
the limits set by the small spin diffusion length. Unfortunately, this is difficult to reconcile
with the coupling between the S-layers observed in the trilayers. Both experiments
point to the presence of triplets in the ESD regime, but with different length scales. The
key here still lies in a better understanding of the inhomogeneous ESD regime, which is not yet available. Just summing the thicknesses of the superconducting and proximized
layer can give a qualitative idea but it is a too simplistic analysis. First of all, the extrapolation
of the $d_{\rm S_{eff}}$ from $H_{\rm c2\parallel}(T)$ is not completely reliable for a proximized system, where $T_{\rm c}$ and therefore $H_{\rm c2}(0)$ are modified. In general, the whole picture can be better described by saying that in these systems the order parameter adjusts itself, such that it can sustain the highest critical field \cite{Nojima}. The outcome, therefore, is the result of the interplay
between parameters such as $\xi_{\rm S}(T)$ and $\xi_{\rm F}$ (or $\xi_{\rm F}^{\rm T}(T)$), with the magnetic field playing a crucial role in determining the magnetic configuration and thus the (possible) triplet generation. 

\section{Conclusions}

To summarize, we showed that equal-spin triplet proximity can be induced in simple Nb/Py/Nb trilayers and Nb/Py bilayers by exploiting the intrinsic magnetic inhomogeneities of Py. The conclusion is indirectly inferred from the DCO observed in the phase diagram $H_{\rm c2 \parallel}(T)$. Indeed, the DCO, observed for both tri- and bilayers (2D--3D transition in the first case, 2D--2D in the latter), cannot be described by a short-range singlet proximity. The crossover appears only for Py thicknesses in the
ESD regime, which seems to provide the optimal degree of inhomogeneous magnetization for the singlet-to-triplet conversion. The interpretation is confirmed by the dependence $T_{\rm c}$ versus $d_{\rm Py}$ for the trilayers, which shows a strong suppression of $T_{\rm c}$ in the ESD regime where the leakage of long-range Cooper pairs is maximum.

\ack

The authors wish to thank S. Bergeret for useful discussions and M. Iavarone for preliminary MFM measurements performed in the early stage of the manuscript. A. G.-S. and J. M. H. thank Universitat de Barcelona for backing their research. This work was funded by the Spanish Government project MAT2011-23698.

{}

\end{document}